\documentclass[referee]{aa}
\usepackage{epsfig}
\begin{document} 
\thesaurus{11(11.13.1; 11.05.2; 09.16.1)} 
\title{Population Synthesis of Pulsars: Magnetic Field Effects}
\author{T. Regimbau, J.A. de Freitas Pacheco} 
\offprints{T. Regimbau}
\institute{Observatoire de la C\^ote d'Azur, B.P. 4229, 06304 Nice Cedex 4, France}
\mail{regimbau@obs-nice.fr, pacheco@obs-nice.fr}
\date{Received date; accepted date} 
\maketitle
\markboth{Regimbau.:~Magnetic field of Pulsars}{}

\begin{abstract}{New results based on methods of population synthesis,
concerning magnetic field effects on the evolution of pulsars are
reported.
The present study confirms that models with timescales for the
magnetic field decay longer than the pulsar lifetime are in better agreement
with data. These new simulations  indicate that the diagram log($P\dot P$) -
log($t_s$) alone cannot be used to test field decay models. The dispersion
of the values of the initial period and magnetic field can explain the observed
behaviour of the data points in such a diagram. The simulations also indicate 
that the statistical properties of anomalous X-ray pulsars
and soft-gamma repeaters (magnetar candidates) are 
compatible with those derived for objects born in the high side tail of 
the magnetic field distribution. The predicted birth rate of neutron stars having field strengths in
excess of  10$^{14}$ G is one object born each 750 yr.}

\end{abstract} 


\section{Introduction}

The magnetic field is probably one of the most important parameters affecting
 the evolution of a pulsar. The field strength fixes the energy loss rate, the 
luminosity in different spectral regions and thus, the observability of 
pulsars. However, since the very first studies on the field evolution 
(Gunn \& Ostriker 1970), the subject has remained highly controversial.
The original statistical study by Gunn \& Ostriker, based on a small
number of objects, led to the conclusion that the magnetic field decays on 
time scales of few Myr, due to Ohmic
dissipation. The analysis of proper motions by Lyne, Anderson \& 
Salter (1982) seems to support that conclusion (see also Narayan \& Ostriker 1990), 
but numerical simulations by Bhattacharya et al. (1992), Hartman et al. (1997)
and Mukherjee \& Kembhavi (1997) suggest that the observed properties of 
the pulsar population are consistent 
with decay times longer than the  pulsar lifetime.

Over the past years there has been increasing recognition that neutron stars 
are born with a considerable diversity of magnetic field strengths and of initial 
rotation periods. In particular, Thompson \& Duncan (1995)  suggested the 
existence of neutron stars, dubbed  ''magnetars'', with 
magnetic field strengths in excess of 10$^{14}$ G.
Observations of soft gamma-ray repeaters  
and  anomalous X-ray pulsars (Kouveliotou et al. 1994, 1998, 1999; Cline et al. 
2000; Vasisht \& Gotthelf 1997; Mereghetti 1999) give some support to the assumption that
 these objects could be associated to highly magnetized neutron stars, if the
braking mechanism is the  ''standard'' magnetic-dipole radiation.  It is worth mentioning
that if the star also loses angular momentum by accelerating a flow 
of relativistic particles, then the derived
magnetic field  should have smaller strengths in comparison with the values
obtained from the canonical model. These sources have typical
X-ray luminosities in the range L$_X \approx$ 10$^{35-36}$ erg/s and 
surprisingly, these values are greater than the
expected magnetic dipole luminosity  I$\Omega\dot \Omega$ for most of the objects. 
A magnetic origin for the X-ray emission
was considered by Thompson \& Duncan (1996), Heyl \& Kulkarni (1998) among 
others, and possible magnetic field decay paths as well as their respective time scale 
were reviewed by Goldreich \& Reisenegger (1992). If magnetic field dissipation
is indeed the energy source of this emission, then typical  decay time scales 
are of the order of  10$^4$ yr  (Colpi, Geppert \& Page 2000).  

 More recently, Urpin \& 
Konenkov (1997) studied the magnetic field evolution of neutron stars, under 
the assumption that the field is initially confined to the crust. According 
to their calculations,
the magnetic field evolution depends on the conductive properties of the 
crust, which are linked to the thermal history of the neutron star. Their 
model computations suggest that the field decays more slowly than the 
classical exponential law resulting from pure Ohmic dissipation. 
They have revisited the diagram originally employed by Gunn \& Ostriker (1970),
in which the quantity $P\dot P$ is plotted against $t_s = P/(2\dot P)$. In 
the standard magnetic dipole model, the former is related to the square of 
the magnetic field strength whereas the latter, to the pulsar age. Urpin \& 
Konenkov interpret this plot as an indication that the magnetic 
field decays, but  recognize
that the observed trend may be caused by the wide range of the period 
derivative values. In fact, Lyne, Ritchings \& 
Smith (1975) conjectured that the observed 
trend in the plane $P\dot P$ - $t_s$ could be explained 
by considering only the dispersion in the observed quantities $P$ and $\dot P$.

In a previous paper (Regimbau \& de Freitas Pacheco 2000, hereafter
paper I) statistical properties of pulsars derived from population
synthesis based on Monte Carlo methods were reported, and the derived
properties of the ''unseen'' objects were used to study their contribution
to the continuous gravitational wave emission 
in the galactic disk. In this work,  we present
results of new simulations, using the up-graded code addressed to the 
following questions: i) is the field decay time scale longer than the pulsar
lifetime as suggested by other simulations ? 
ii) are the anomalous X-ray pulsars
(AXP) and the soft gamma-ray repeaters (SGR) a distinct class of  neutron stars ?
In other words, if a high  magnetic field is responsible for the observed
characteristics, are these objects simply associated to neutron stars born 
in the high side tail of the field distribution ?
iii) Can the observed trend in the plane $P\dot P$ - $t_s$ be explained only
by dispersion of the initial values of the rotation period and  magnetic
field ?  The plan of this paper is the following: in section 2 we describe briefly 
the model in which the numerical code is based on, in section 3 we report
the main results of this work and finally, in section 4 we discuss and 
present our main conclusions.  

\section{The Model}

A detailed presentation of our model was already reported in paper I and here,
for the sake of completeness, only the principal  aspects are reviewed. 

In our code, pulsars are generated at a constant rate with initial periods and magnetic 
fields distributed according to Gaussian and  log-normal probability 
functions respectively. Their initial
positions in the Galaxy are established by assuming exponential 
distributions, having a scale of height of 100 pc along the z-axis and 
2.3 kpc along the radial direction. A scenario excluding recent pulsar
formation in the inner galactic bulge (1 kpc around the galactic center)
was also considered, but the final results are not considerably affected by this
assumption. The effect of an initial random kick 
velocity on the orbital motion is also taken into account.   In
our simulations, the transversal and radial (coincident with the direction of the orbital motion) 
components of the kick velocity are assumed to have a Gaussian distribution (Lorimer 1993)
with velocity dispersions of 140 km/s and 100 km/s respectively, corresponding to an average
kick velocity of 170 km/s. Simulations with average velocities of 300 km/s (Hansen \& Phinney
1997) and 450 km/s (Lorimer \& Lyne 1994) were also performed without any appreciable
modification in the results. Simulations with high kick velocities produce a  significant number
of objects which escape from the Galaxy, modifying slightly the pulsar birthrate. 

In order to model the evolution of  the rotation period, two scenarios were 
considered. In the first, the standard magnetic dipole model was modified to 
permit a continuous increase of the angle between the spin and  the magnetic 
dipole axes, but remaining constant the field intensity (model A). In the second, the
magnetic field was allowed to decay exponentially (model B). 
The migration of the magnetic dipole from an initial arbitrary angle to an 
orthogonal position with respect to
the spin axis is a consequence of the scenario explored by Link, Franco \& 
Epstein (1998) and Epstein \& Link (2000), where starquakes in 
spinning-down pulsars
may push matter toward the poles, causing an increasing misalignment of 
those axes. This possibility is quite attractive, since during the migration 
phase braking indices smaller than three, as found in some young pulsars, 
can be obtained  (see paper I) and it could also be a
possible explanation for the increasing spin-down rates observed in some
pulsars and SGRs, as we shall see later. In this case, the period evolution is 
given by
\begin{equation}
P = P_0\lbrack 1 + {{t}\over{\tau_0}} - n_0{{t_{\alpha}}\over{\tau_0}}(1 -
e^{-t/t_{\alpha}})\rbrack^{1/2}
\end{equation}
where $P_0$ is the initial period, $\tau_0$ 
= ${{3}\over{4\pi^2}}{{Ic^3P_0^2}\over{B^2R^6}}$
is the magnetic braking time scale and $t_{\alpha}$  is the magnetic
dipole migration time scale. The
parameter $n_0$ corresponds to the initial angle $\alpha$ between the spin 
and the magnetic axes such as $\alpha$ = arcos($\surd n_0$).  Tauris \& Manchester (1998) from the analysis of polarization data,  found a tendency for the
magnetic axis to align with the rotational axis on a timescale of about 10$^7$ yr.  In fact, they
assumed a time evolution for the angle $\alpha$ of the same form as that suggested by Jones
(1976), namely,
\begin{equation}
sin \alpha(t) = sin\alpha_o e^{-t/t_a}
\end{equation}
In this situation, the solution for the period evolution is formally identical that obtained when
one assumes an exponential decay for the magnetic field (case B). Thus, the results of
case B (discussed below) concern also the scenario envisaged by Tauris \& Manchester
(1998).  It should be emphasized, as those authors did, that
such a scenario implies a decreasing magnetic torque and, as a consequence,
a braking index greater than three, in disagreement with observations.

The detectability of pulsars is affected by: a) the fact that their radio emission
is not isotropic, b) pulse broadening due to instrumental effects and
c) dispersion and scattering through the interstellar medium.  All these effects 
are included in our simulations (see paper I for details).

Our data are a culled sample of  491 pulsars extracted from the up-graded catalog of
Taylor, Manchester \& Lyne (1993), where only single objects supposed to be
originated from population I stars and with all required parameters measured
were considered (The word 'single' here does not exclude the possibility that
the progenitor evolved in a binary system, which was disrupted at the moment of
the explosion, imparting momentum to the newly born neutron star). 
Pulsars not included in any of the original surveys were
also discarded. Since this catalog is a compilation of different surveys
covering specific sky areas and having a well defined sensitivity, the objects
were distributed in sub-classes, according to the original surveys in which
they were first detected. In the present study, pulsars detected by the
76m radio telescope at Jodrell Bank
were also included in our sample. The main characteristics of this
instrument to be added to table 1 of paper I are: T$_r$ = 50 K, $\tau_{sam}$ = 0.6 ms,
C$_{DM}$ = 0.016 ms.cm$^3$.pc$^{-1}$ and A$_0$ = 0.19 mJy/K.

For each pulsar generated, our code follows its evolution in order to compute
the present $P$, $\dot P$ values and galactic coordinates defining the
survey (or surveys) included in the general catalog, that covered such a
region of the sky. The next step is the application of  ''filters'' 
defining the detectability of the object, as mentioned above. 
Our procedure insures that the simulated catalog will have
objects detected by different surveys distributed in the same proportion
as in the global catalog. The numerical experiments generate objects
according to these prescriptions until the simulated sample be equal to the
data sample. For a given run, defined by a set of initial parameters,
the number of experiments is comparable to the number of objects in the actual
data sample and the final result is a suitable average, in order to avoid
statistical fluctuations. For each experiment so defined, we 
compare the resulting distributions with those derived from present data.
The values of the input parameters were optimized, by controlling the fit
quality through $\chi^2$ and Kolmogorov-Smirnov tests.

\section{Results}

Until a recent past, the general belief was that most pulsars were born with parameters
similar to those of the youngest known objects: the Crab ($P$ = 33 ms) and
PSR 0540-69 ($P$ = 50 ms) are well known examples. Different observational facts
suggest that most pulsars are born with periods $P \geq$ 100 ms (see, for instance,
Bhattacharya 1990). Indeed, fifteen young pulsars with ages less than 50 Kyr listed
in the updated Taylor, Manchester \& Lyne (1993) catalog have an average 
period equal to 200 ms.

The results of the present simulations compared with a data sample larger
than that used in paper I, confirm that the average initial rotation period  of pulsars
is $<P_0>$ = 290 ms with a dispersion of 100 ms (model A). It should be 
emphasized that these values give the best representation of  the observed
period distribution even when the magnetic field decay
is included (model B). In figure 1 the simulated period distribution for
both scenarios is compared with the present data and 
the optimized input parameters for these models are given in table 1.
In columns two and three, the parameters defining the initial (Gaussian) 
distribution of periods and (log-normal) magnetic field braking time scale (average and 
dispersion values) are respectively given; in columns four and five
are respectively indicated the pulsar lifetime and the magnetic field
decay time scale.
Notice that model B requires a higher magnetic braking time scale $\tau_0$
and a smaller dispersion of this quantity. This is a consequence of the
decreasing magnetic torque due to the field decay, requiring a longer
time interval to produce the same deceleration. Concerning the dipole migration
time scale $t_{\alpha}$, the new simulations simply confirm the conclusions
of paper I.  It is worth mentioning that three magnetar candidates 1E 1048-5937
(Mereghetti 1995), SGR 1900+14 (Kouveliotou et al. 1999) and  1RX J1708-40 (Kaspi,
 Lackey \& Chakrabarty 2000), with
indicative ages $t_s < t_{\alpha} \approx 10^4 yr$ (see table 2 below), have 
undergone episodes in which the spin-down rate has almost doubled. The
resulting enhanced magnetic torque is consistent with the
migration scenario originally developed by Link, Franco \& Epstein (1998).

\begin{table*}
\caption[1]{Optimized Model Parameters}
\begin{flushleft}
\begin{tabular}{lcccccccccc}
\noalign{\smallskip}
\hline
\noalign{\smallskip}
Model& P$_0 \pm \sigma_{P_0}$ (ms)& $ln(\tau_0) \pm \sigma_{ln(\tau_0)}$&
t$_{max}$ (Myr) &t$_D$ (Myr)\\
\noalign{\smallskip}
\hline
\noalign{\smallskip}
A & 290$\pm$100 & 9.0$\pm$3.6&25& $\infty$ \\
B & 290$\pm$100 & 10.6$\pm$2.6&25&15\\
\noalign{\smallskip}
\hline
\end{tabular}
\end{flushleft}
\end{table*}

However, the situation is completely different concerning the period derivative
distribution. The inclusion of the field decay displaces the maximum
of the distribution toward lower values, since the decreasing magnetic torque 
 weakens the pulsar deceleration,  producing more
objects with smaller period derivatives.
Figure 2 compares the simulated
period derivative distribution derived  for models A and B with actual
data where the mentioned effect can clearly be seen. If the initial parameters 
of model B are modified, in particular the magnetic field, in order to 
improve the fit quality, then a good representation of the period 
distribution is no longer obtained. Hartman et al. (1997) had similar
difficulties in their simulations, since they were not able to simultaneously
reproduce the distributions of periods and magnetic fields using models
with short decay time scales.

The plane $P\dot P$ - $t_s$, which is equivalent to a plot
of the square of the magnetic field  versus the indicative age of the pulsar, has been
used by several authors to test effects of the field decay (Gunn \& Ostriker 
1970; Urpin \& Konenkov 1997). Our simulations indicate that even without
including magnetic dissipation, the observed data trend  on that plane 
can be reproduced quite well (figure 3). Model B does not improve the results
and, in fact, a combined
statistical test of the distribution of points along both axes favors model A.
The distribution of the points can be characterized by the average slope and 
the scatter amplitude about  the mean locus. The simulations indicate that these
parameters  depend on the 
dispersion of the  magnetic field (mainly) and on the dispersion
of the initial rotation period, in agreement
with the supposition by Lyne, Ritchings \& Smith (1975). The importance
of the dispersion of the initial field value was checked by performing
simulations under the assumption that all pulsars are born with the same
magnetic field.  In this case, the dispersion of the
initial periods only is not enough to explain the observed trend. If 
the dispersion of the initial period distribution is increased, the simulated
data are more scattered, but the slope cannot be reproduced. In this 
case, models including magnetic 
dissipation improve the results but always with a poorest fit quality in comparison
with model A.

Another interesting point concerns the comparison between the simulated
magnetic field distribution of the observed and of the real (''unseen'')
pulsar populations. These
distributions are displayed in figures 4 and 5 for models A and B respectively.
The average field derived for the observed population is 2.5 $\times$ 10$^{12}$ G for
model A and 1.6 $\times$ 10$^{12}$ G for model B, in good agreement with the
values obtained from the canonical model and data, but the fit quality for model
A is better.
As already noticed by other authors
(see, for instance, Bhattacharya et al. 1992) the average magnetic field of the
real population is higher than that of the observed one, and this tendency can clearly
be seen in figures 4 and 5. In the case of model B, the average field of the unseen
pulsars is about three times higher, whereas model A generates a population with
an average field one order of magnitude higher than that of the detected pulsars.
Such an effect can be explained by the following reasons: firstly, pulsars generated 
at birth in the high side tail of the magnetic field distribution evolve rapidly to long
periods. Consequently, their radio-luminosity decreases quite fastly and the
detection probability becomes rather small. Secondly, according to the beaming
model by Biggs (1990), adopted in the present simulations, as period increases
the fraction of the sky covered by the pulsar emission cone decreases, reducing
also the detection probability of those high field or long period objects. The
latter case is very well illustrated by the radio pulsar J2144-3933 
(Young, Manchester \& Johnston 1999), which has a period of 8.51 s and a very
narrow pulse profile, with a half intensity of about one degree of longitude.

The statistics of the high field (B $> 10^{14}$ G ) population are 
interesting to mention in some detail, in order to verify if they are compatible 
with the observed properties
of AXPs and of SGRs. We report here essentially the results derived from model
A, since from our point of view, they describe more adequately the high field population.

The pulsar birth rate derived from model A is one object born each 170 yr. Since
the high field objects correspond to about 23\% of the total population, their
birth rate corresponds to about one object born each 750 yr.

From our simulations, the number of high field objects with rotation period less 
than P is approximately given by  $ N(<P) \approx 0.85P^{1.85}$. This equation
predicts that the number of high field objects in the Galaxy with P$\leq$ 11 s is
about 72. Most of these objects have magnetic field strengths in the range
14.0 $<$ log B $<$ 15.5 and no higher field objects are expected to be detected
in this period interval, since their number becomes appreciable only for periods
longer than 130 s.  The present detected number of AXPs and SGRs 
with P $\leq $ 11 s is about a dozen, suggesting that only a small 
fraction  ($\approx 15\%$) of these objects was already discovered.  
 These numbers must be taken with a grain of salt, since the selection effects on the
detection of  SGRs and AXPs are rather different of those present in pulsar surveys,
simulated in our numerical experiments.

Seven objects suspected to be magnetars have their period derivative measured.
For these objects, the indicative age ${t_s} = P/(2\dot P)$ can be
derived  and a comparison with simulated data is possible. However, for
young objects $t_s$ may overestimate the true age, depending on the initial
period and on the initial field strength. Therefore, the simulated objects were classed
into different bins of age and magnetic field, and their average properties
like period and indicative age were computed after 10$^5$ runs. In figure
6, the mean period is plotted against the mean indicative age and each
curve corresponds to a given  magnetic field interval. The available
data (summarized in table 2) are consistent with the statistical properties
of the simulated population, supporting the thesis that AXPs and SGRs
are  neutron stars born in the high side tail of the  magnetic field distribution.

\begin{table*}
\caption[2]{Magnetar Candidades}
\begin{flushleft}
\begin{tabular}{lcccccccccc}
\noalign{\smallskip}
\hline
\noalign{\smallskip}
Object& P (s)& log($\dot P$)& log($t_s$) (yr)& references \\
\noalign{\smallskip}
\hline
\noalign{\smallskip}
SGR 1806-20& 7.47 & -10.081 & 3.176 & Kouveliotou et al. 1998\\
1E 1841-045& 11.77 & -10.328&3.620& Vasisht and Gotthelf 1997\\
1E 1048-59& 6.45& -10.657: & 3.689 & Oosterbroek et al. 1998\\
4U 0142+61& 8.68& -11.638& 4.800& Mereghetti and Stella 1995\\
1E 2259+58& 6.98& -12.136& 5.202& Mereghetti and Stella 1995\\
SGR 1900+14& 5.16& -9.958& 2.893& Kouveliotou et al. 1999\\
1RX J1708-40& 11 & -10.632& 3.895& Israel et al. 1999\\
\noalign{\smallskip}
\hline
\end{tabular}
\end{flushleft}
\end{table*}

\section{Discussion and Conclusions}

Simulations performed by Bhattacharya et al. (1992), Hartman et al. (1997)
and Mukherjee \& Kembhavi (1997)
lead those authors to conclude that no significant field decay occurs (t$_D >$ 100-160 
Myr) during the active life time of the pulsar. The present  simulations essentially
confirm these results, since models without field decay are those which give the best
representation  of data. When the field decay is included in the pulsar evolution, we were
unable to find a set of input parameters, characterizing
the initial period and magnetic field distributions, which give a simultaneous acceptable
representation  of  the observed period, first derivative and magnetic field distributions.

Our simulations indicate that the diagram  $P\dot P$ - $t_s$ alone cannot be used to
test the magnetic field decay without ambiguity. The dispersion of the values 
of the initial field and rotation period can explain by themselves the observed 
trend of the data points, namely, the  average slope
and scattering amplitude. Concerning the interpretation of that diagram, it is worth mentioning 
the following point: let us assume for a while that the field dissipation controls the
distribution of the data points. Then, the best fit gives a relation of the form
\begin{equation}
log(P\dot P) = a + b.log(t_s)
\end{equation}
where a = -3.86 and b = -0.78 (c.g.s. units). This equation can immediately 
be integrated, resulting for the period evolution
\begin{equation}
P \propto t^{(1+b)/2} \propto t^{0.11}
\end{equation}
This equation is equivalent to have assumed a power law for the magnetic
field decay (B $\propto t^{b/2} \propto t^{-0.39}$)  in the equation of motion.
As a consequence of field decay, the pulsar deceleration is smaller than
that predicted by the standard model and, in this case, the expected
braking index  N = ${{\ddot \Omega \Omega}\over{\dot \Omega^2}}$  is
equal to ${{(3+b)}\over{(1+b)}} \approx 10$, almost the triple of the
canonical value (N = 3). Four young pulsars with measured  braking indices
have N $<$ 3 and, according to the scenario discussed in
paper I, these objects are probably in a phase where the  magnetic dipole is
still migrating. A relatively high magnetic field (B$\approx 4 \times
10^{13}$ G) pulsar (J1119-6127) recently discovered (Kaspi et al. 1999) has
a normal braking index (N = 3), suggesting that no direct evidence of
the magnetic field decay has been observed until now in the period evolution.

The expected number of high field  (B $> 10^{14}$ G) pulsars resulting from
model A is about 23\% of the total population and about 72 objects members
of this class, having periods less than 11 s are expected to be present in the
Galaxy. The number of these objects predicted by model B is rather modest, of
the order of  5-6. Since a dozen of 
magnetar candidates have already been discovered, this is another argument in
favor of model A,  which also predicts for the high field population
relations between the mean period and mean indicative age
compatible with the present available data on these objects. These statistical  properties
are not in contradiction with the interpretation that magnetars are 
objects born in the high side tail of the magnetic field distribution.
 
If this scenario is correct, why AXPs are radio quiet ?  Pair production
by the mechanism $\gamma \rightarrow e^+e^-$ is a necessary condition
invoked in most pulsar emission models (Sturrock 1971). However, for fields in
excess of the quantum critical field $B_{cr}$ = ${{m^2c^3}\over{e\hbar}}$
= 4.41$\times 10^{13}$ G, Baring \& Harding (1998) argued that photon splitting
($\gamma \rightarrow \gamma\gamma$) becomes dominant, suppressing
pair creation and the magnetosphere required for radio emission. The
discovery of  radio pulsars with magnetic fields comparable to
the critical value (Kaspi et al. 1999; Camilo et al. 2000) difficults such a simple explanation.
The photon splitting mechanism alone is not able to explain why two young
pulsars ($t_s < $ 2000 yr), J1119-6127 (Kaspi et al. 1999) and J1846-0258
(Gotthelf et al. 2000) having quite similar
periods ( P = 0.407 s and 0.323 s respectively) and magnetic fields ( B =
0.93B$_{cr}$ and 1.13B$_{cr}$ respectively) behave so dissimilarly. The
former is a radio pulsar whereas the latter is a X-ray pulsar, associated with
the supernova remnant Kes 75. No radio emission has been observed from
J1846-0258, but the object is probably located inside the central core of the SNR, a
synchrotron plerion similar to the Crab, rendering difficult such a detection.
More recently, Zhang \& Harding (2000) suggested that besides photon
splitting, the orientation of the magnetic dipole with respect to the spin axis may
also play an important role to establish the existence or not of radio emission, but
this question is still waiting for a more satisfactory answer.

The energy source of the observed X-rays  is another problem to be faced if
such an emission is not fed by field dissipation. A possible issue was considered
by Heyl \& Hernquist (1997), who showed that for B$> 10^{12}$ G, the quantization
of the electron energies enhances the conductivity along the field lines. As
a consequence, there is a net increase in the heat flux and high field objects
may have effective temperatures up to 40\%  larger than low field neutron stars.
Thus, enhanced cooling  induced by a strong magnetic field could  be a possible
explanation for the X-ray emission observed in AXPs.    

In conclusion, the magnetic field strength seems to play a major role in the
evolution of young neutron stars. Objects born with B$\leq 10^{13}$ G evolve
as radio-loud pulsars, since curvature radiation produces a pair magnetosphere
required for  most emission models. In higher field objects (B$> 10^{14}$ G), the
quantum process of photon splitting acts as a quenching mechanism for pairs,
suppressing radio emission. In these stars, the high field also enhances the cooling and 
produces higher luminosities as well as temperature variations across the
surface, which can explain the observed pulsed emission of AXPs. Three
objects are presently known in the transition region $10^{13} < B < 10^{14}$ G
and, as mentioned above, the understanding of their properties requires a
considerable improvement of the existing models.

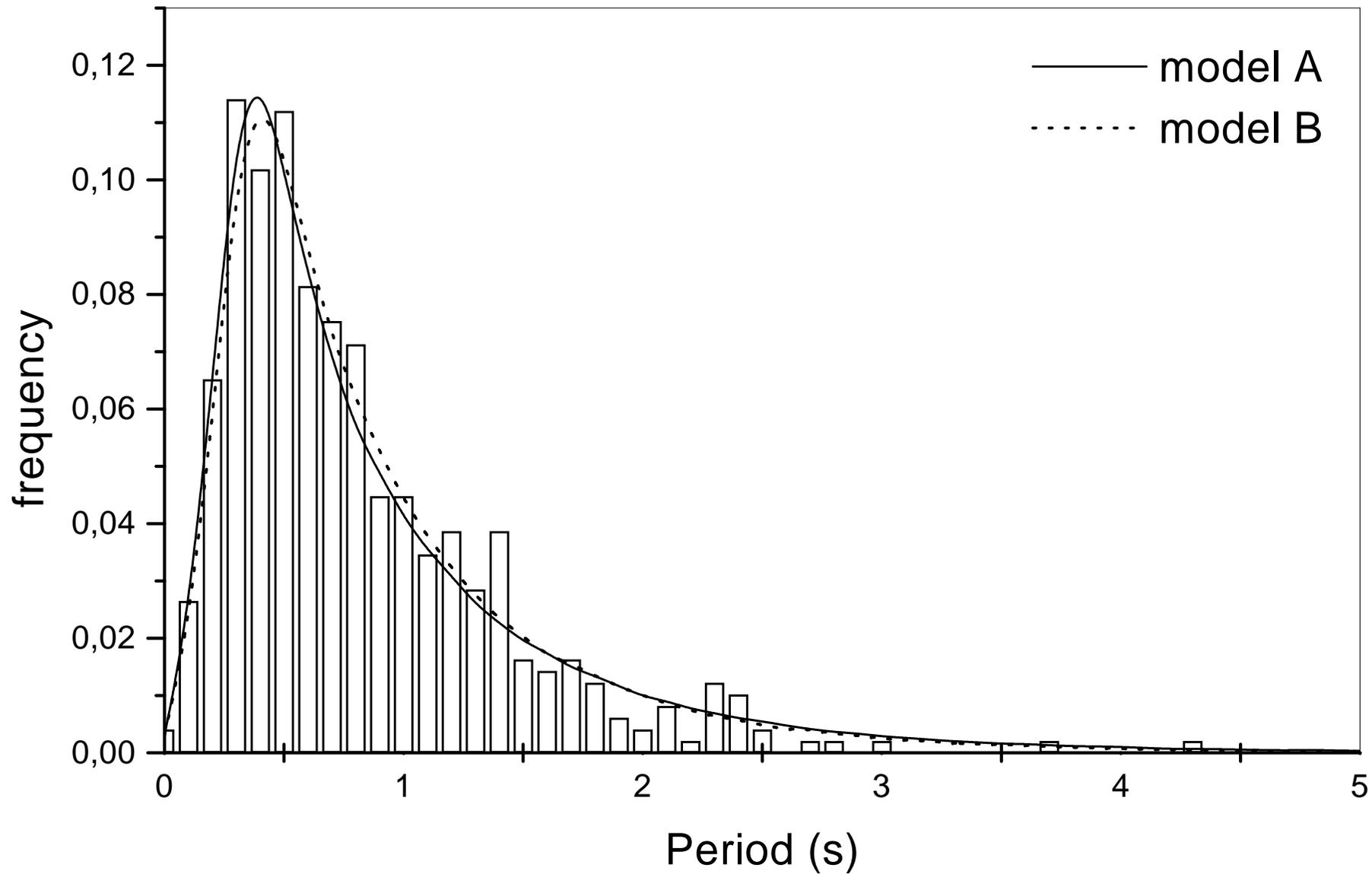
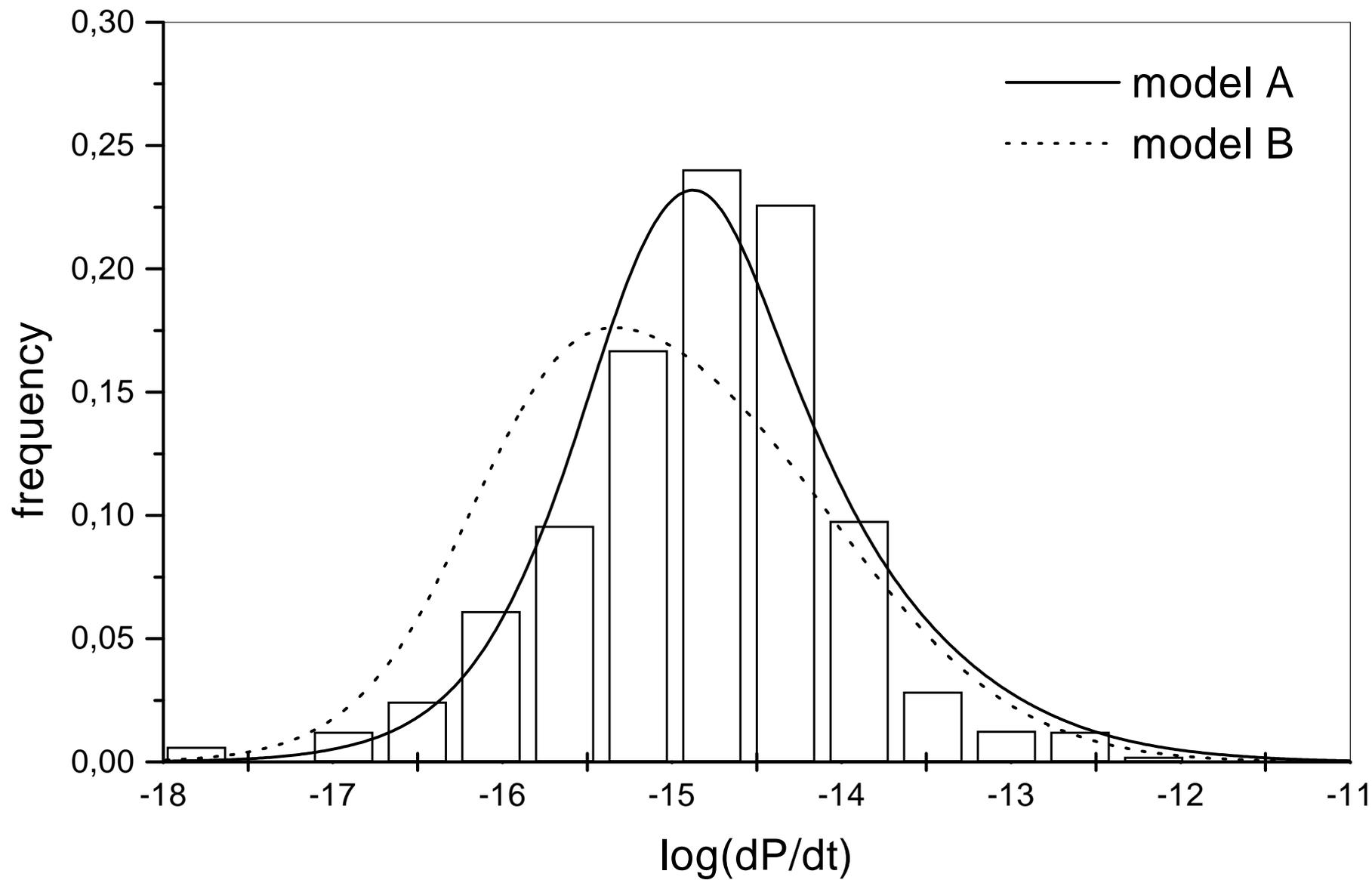
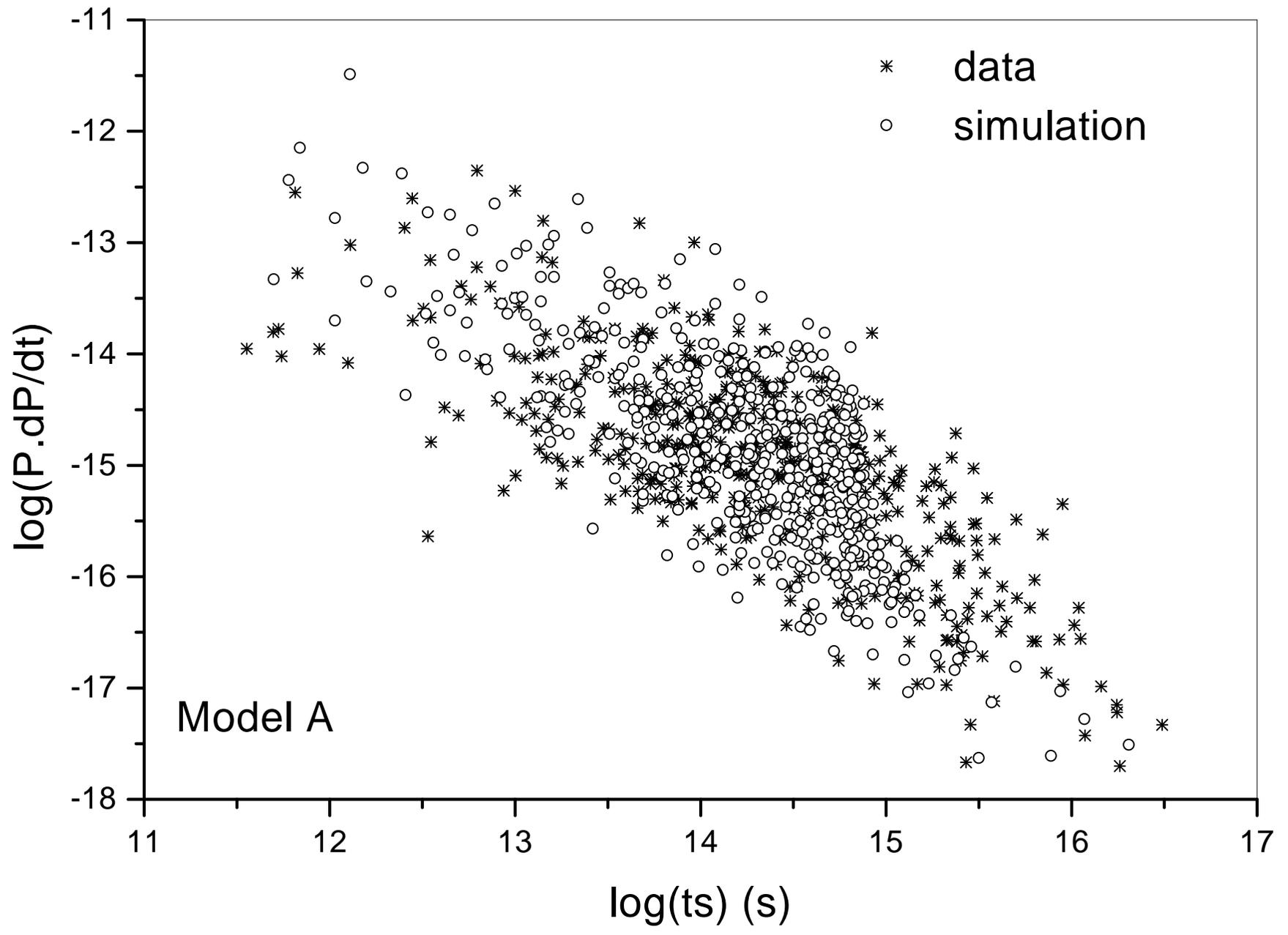
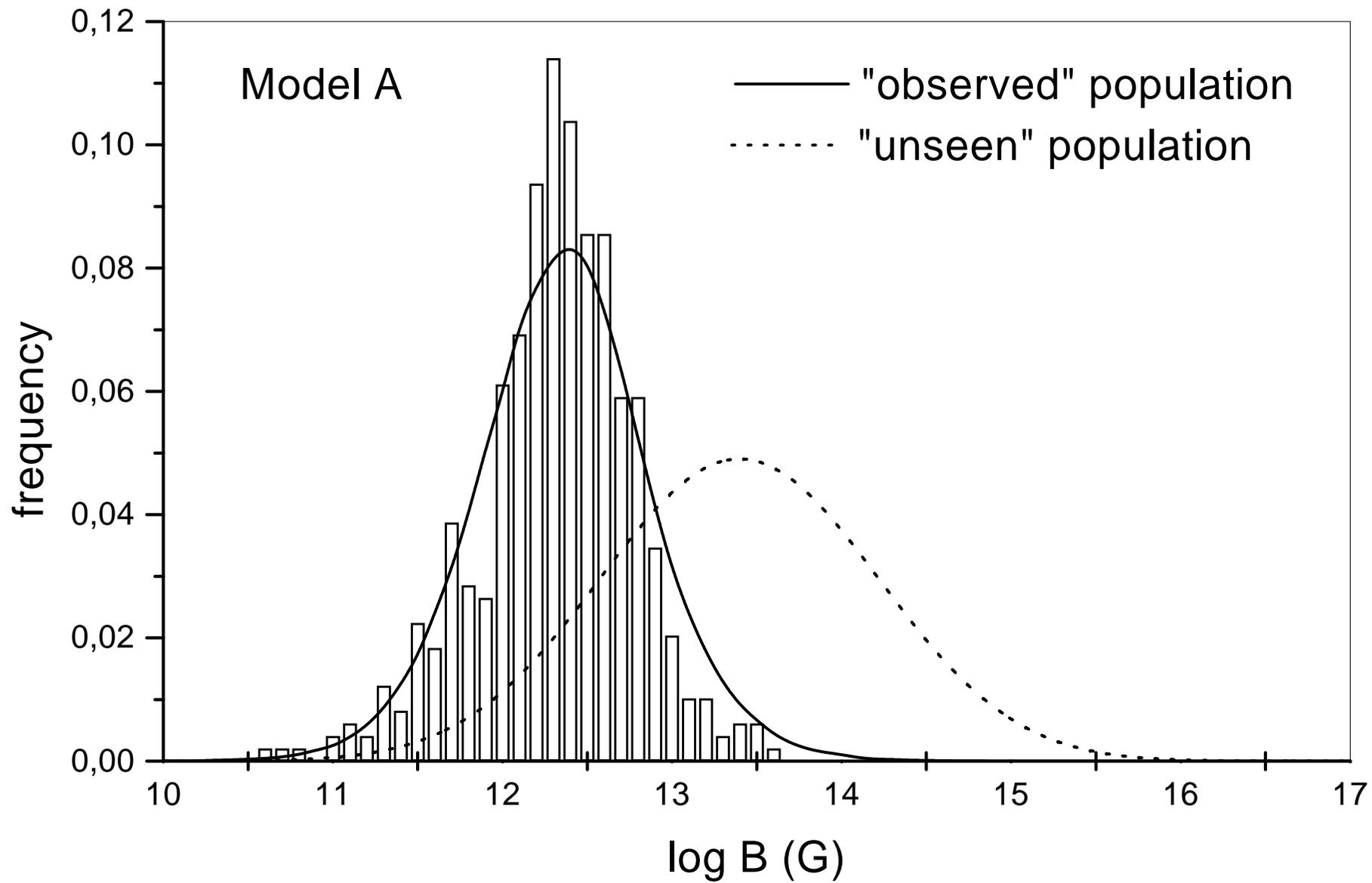
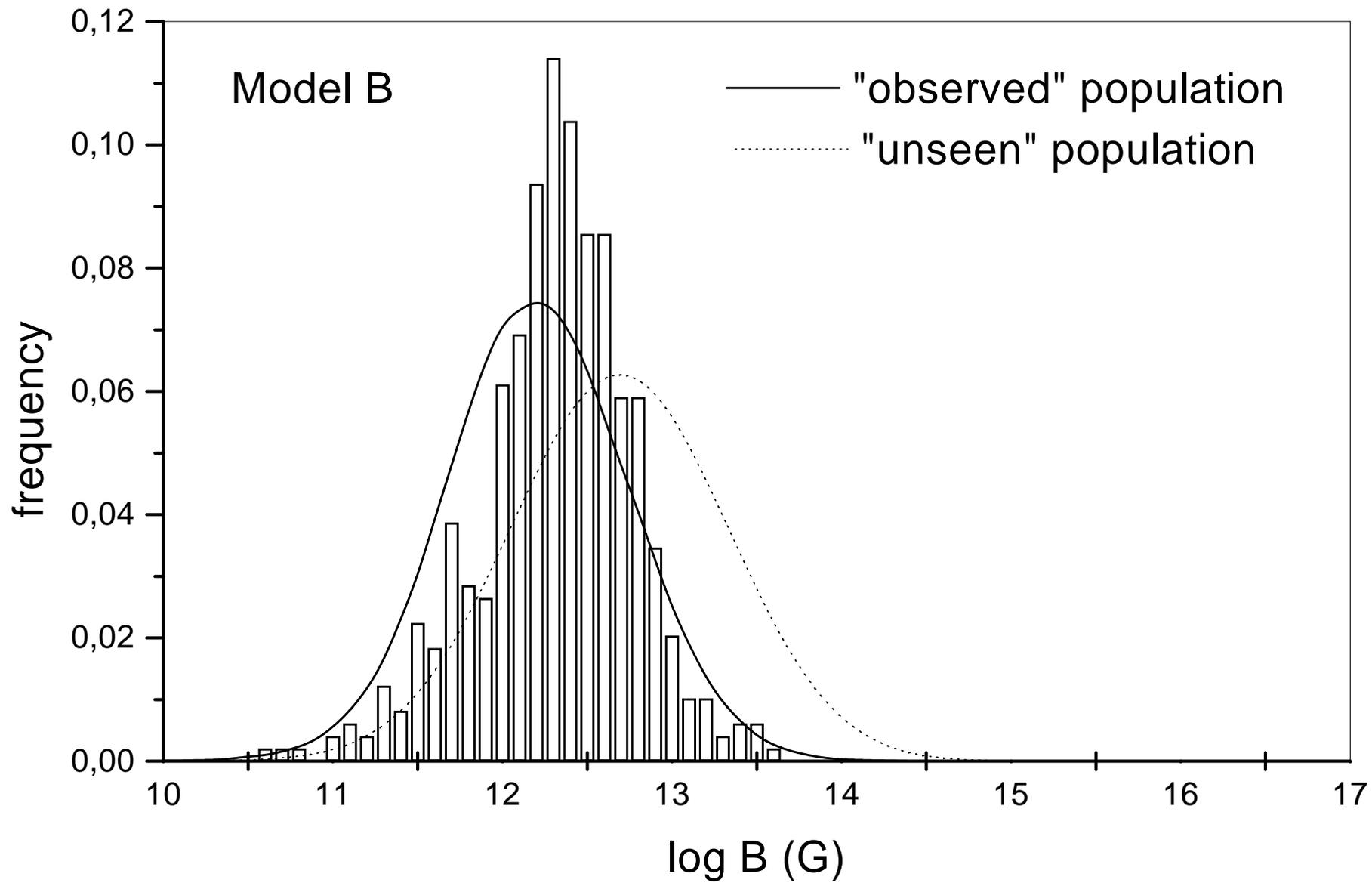
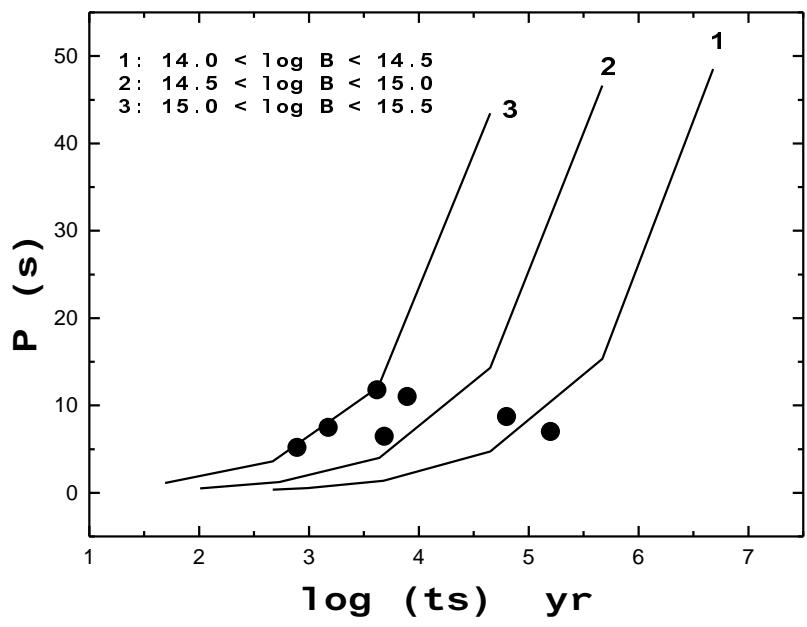
\end{document}